# SYNTHESIS AND PHYSICAL PROPERTIES OF $FeSe_{1/2}Te_{1/2}$ SUPERCONDUCTOR

V.P.S. Awana*, Anand Pal, Arpita Vajpayee, Monika Mudgel, and H. Kishan

National Physical Laboratory, Dr. K.S. Krishnan Marg, New Delhi 110012, India

Mushahid Husain

Department of Physics, Jamia Millia Islamia, New Delhi-110025, India

R. Zeng*

ISEM, University of Wollongong, NSW 2522, Australia

S.Yu[1,2], Y. F. Guo[1,3], Y.G. Shi[1,3], K. Yamaura[1,2], and E. Takayama-Muromachi[1,2,3]

[1]JST, Transformative Research-Project on Iron Pnictides (TRIP), Tsukuba, Ibaraki 305-0044, Japan

[2]Superconducting Materials Center, National Institute for Materials Science, 1-1 Namiki, Tsukuba, Ibaraki 305-0044, Japan

[3]International Center for Materials Nanoarchitectonics (MANA), National Institute for Materials Science, Tsukuba, Ibaraki 305-0044, Japan

* Presenting and Corresponding Author

Dr. V.P.S. Awana

National Physical Laboratory, Dr. K.S. Krishnan Marg, New Delhi-110012, India

Fax No. 0091-11-45609310: Phone No. 0091-11-45608329

e-mail-awana@mail.nplindia.ernet.in: www.freewebs.com/vpsawana/

**Abstract**

One of the most important properties of very recently reported FeSe based superconductors is the robustness of their superconductivity under applied magnetic field. The synthesis and control of superconductivity in FeSe based compounds is rather a difficult task. Synthesis and physical property characterization for optimized superconductivity of $FeSe_{1/2}Te_{1/2}$ at 13 K is reported here. The compound crystallized in a tetragonal structure with lattice parameters a = 3.8008(10) and c = 6.0187 (15) Å. Magnetization measurements indicated bulk superconductivity with lower critical field ($H_{c1}$) of around 180 Oe. By applying Ginzburg Landau (GL) theory, the $H_{c2}(0)$ value is estimated to be ~1840 kOe for the 90% of resistive transition. A heat capacity measurement revealed bulk superconductivity by a hump at $T_c$ near 13 K, and an expected decrease of the same was observed under an applied magnetic field.

## 1. Introduction

The recent discovery of iron-based chalcogenide [1-3] and the iron-based pnictide [4-6] superconductors has generated a great deal of interest in the scientific community. The parent compounds of the FeAs based superconductors do not exhibit superconductivity but possess stripe type commensurate antiferromagnetic spin order accompanied by a structural transition at around 150 K [7-8]. The superconductivity was obtained by either fluorine doping at the oxygen site or by the deficiency of oxygen ions in the system [9-11]. The highest $T_c$ = 55 K has been reported in $SmFeAS(O_{1-x}F_x)$ [12]. Very recently, another iron based superconductor FeSe was reported with a $T_c$ of 8 K [1]. FeSe is the simplest-structured among the iron-based superconductors. The $T_c$ of FeSe is increased up to 15 K by partial substitution of Te or S for Se and it is also increased up to ~27 K by applying hydrostatic pressure of 1.48 GPa [3,13-17]. The effect of pressure on superconductivity i.e. $dT_c/dP$ is around 9.1 K/GPa, which is four times more than for any other known superconductor till date. Very high $dT_c/dP$ of FeSe warrants large scope for increasing its $T_c$ by chemical pressure through on site substitutions. Garbarino et. al found that under compression, FeSe transforms from the original tetragonal structure to an orthorhombic high pressure phase above 12 GPa. The high-pressure orthorhombic phase has a higher $T_c$ reaching 34 K at 22 GPa [18].

Here, we report synthesis and physical properties of $FeSe_{1/2}Te_{1/2}$. The structural, magneto-transport and heat capacity measurement are reported for title compound.

## 2. Experimental

A Polycrystalline sample of title compound was synthesized by the solid-state reaction route. The stoichiometric ratio of high purity (> 3N) Fe, Se and Te were ground, palletized and encapsulated in an evacuated quartz tube. The encapsulated tube was then heated at 750$^o$C for over 24 hours and slowly cooled to room temperature. The sintered pellets were again ground, pelletized in a rectangular shape, sealed in an evacuated quartz tube, and re-sintered at 750$^o$C for 12 hours. Unlike the highly reactive RE (rare earths) and alkaline metals (Sr, Ba, Ca), the Fe, Se and Te were weighed and mixed in open with inhaling (mask) and skin touch (gloves) precautions. The x-ray diffraction patterns of the samples were obtained with the help of a Rigaku diffractometer. The resistivity measurements were recorded for temperatures down to 4.2 K via a four-probe method. The temperature dependence of DC magnetization of the $FeSe_{1/2}Te_{1/2}$ sample was analyzed using a SQUID magnetometer (Quantum Design) both in zero field and field-cooled configurations. Heat capacity measurements were carried out on a Quantum design PPMS.

## 3. Results and Discussion

The x-ray diffraction pattern of $FeSe_{1/2}Te_{1/2}$ compound and its Rietveld refinement are shown in Fig. 1. All the peaks are well indexed using a space group of P4/*nmm*, and no impurity phases are found. Fe is located at coordinate position (¾, ¼, 0) and Se/Te at (¼, ¼, 0.2778) in $FeSe_{1/2}Te_{1/2}$ compound. The compound crystallizes in a tetragonal structure and no secondary phase is observed. Lattice parameters are found to be a = 3.8008(10) and c= 6.0187 (15) Å. The lattice parameters are in good agreement with previously reported literature [13, 18].

The magnetic susceptibility at 10 Oe is shown in Fig. 2 for both ZFC and FC measuring conditions. Sample shows clear superconducting diamagnetic response below this onset temperature $T_c^{dia}$ i.e. below 13 K. This value is in confirmation with earlier reports [19-21]. This figure confirms the bulk superconductivity in the present sample. The hysteresis loops (M-H) measured at temperatures 2, 5 and 8 K are shown in lower inset of Fig. 2. The upper inset of Fig.2 shows only the first quadrant of the M(H) loop at 2, 5 and 8 K. The M(H) plot inverts from around 180 Oe at 2 K. For higher T of 5 and 8 K, obviously the inversion field and hence $H_{c1}$ decreases. This shows that the lower critical field ($H_{c1}$) of the studied $FeSe_{1/2}Te_{1/2}$ sample is around 180 Oe at 2K. The lower inset depicts wide open M(H) plots in all four quadrants at 2, 5 and 8 K. The opening of the M(H) loops decrease with increase in temperature. Further the positive moment of the Fe magnetic ion is riding over the diamagnetic signal and hence results in some positive background.

Figure 3 shows the temperature dependence of the resistivity ρ(T) of the $FeSe_{1/2}Te_{1/2}$ sample. This figure depicts that studied sample has somewhat semiconducting behavior above 150 K, below this temperature it shows the metallic behavior in the normal state. Mizuguchi et. al also got the semiconducting type behavior for Te substituted FeSe compounds [13]. With decrease in temperature below 14 K, the resistivity of the sample starts vanishing abruptly due to occurrence of superconductivity. The onset and the zero resistivity temperatures of the sample are determined to be 13.5 K and 11.5 K, respectively.

In order to determine the upper critical field of the sample ρ(T) curves are measured under different magnetic fields up to 90 kOe. The superconducting transition zone of ρ(T)H measurements is shown in lower inset of Fig. 3. The superconducting transition temperature decreases with an increase in applied magnetic field with a rate of decrement i.e. $dT_c/dH$ ~ 0.03K/kOe. Such a low value of $dT_c/dH$ makes this compound superior against the Nb based BCS type, HTSc, Borides or even the recently invented FeAs superconductors.

The upper critical field is determined using different criterion of $H_{c2}=H$ at which $\rho=90\%\rho_N$ or $50\%\rho_N$ or $10\%\rho_N$ where $\rho_N$ is the normal resistivity or resistivity at about 15 K. The $H_{c2}$ variation with temperature is shown in upper inset of Fig. 3. To determine $H_{c2}(0)$ value, we applied Ginzburg landau (GL) theory. The GL equation is:

$$H_{c2}(T)=H_{c2}(0)*(1-t^2)/(1+t^2)$$

Where, $t=T/T_c$ is the reduced temperature [22]. The fitting of experimental data is done according to the above equation, which not only determines the $H_{c2}$ value at zero Kelvin [$H_{c2}(0)$] but also determines the temperature dependence of critical field for the whole temperature range. $H_{c2}(10\%)$, $H_{c2}(50\%)$ and $H_{c2}(90\%)$ are estimated to be 608, 880 and 1842 kOe respectively at 0 K. These values of upper critical field are comparable with the reported in literature [17,23,24]. Yadav et al. [24] estimated the values of $H_{c2}(10\%)$, $H_{c2}(50\%)$ and $H_{c2}(90\%)$ as 690, 880 and 1840 kOe respectively for $FeTe_{0.60}Se_{0.40}$ sample.

In order to study the nature of magnetic anomalies we further performed specific heat measurement under a magnetic field of 0 and 70 kOe, as shown in Fig. 4. The specific heat in zero-field displays a small hump at $T_c \approx 13$ K. This is in agreement with literature [19]. After applying the magnetic field of 70 kOe the small hump is observed at the same temperature but with slightly decreased magnitude. This is obvious for the case of a superconducting transition. Generally, specific heat of a superconductor decreases with applied. The enlarged view of this anomaly present in $C_p/T$ vs $T^2$ is shown in inset of Fig. 4. There is no report available on heat capacity measurement under applied magnetic field in this temperature range; in Ref. 19 Sales et al. reported the $C_p(T)$ plot only at zero field. Very recently [25], in-commensurate magnetic excitations are reported for $FeSe_{1/2}Te_{1/2}$ just below its $T_c$, which could also contribute to $C_p(T)H$ anomaly.

**Conclusions**

In summary, we have studied the structural, electronic transport and magnetic properties of $FeSe_{1/2}Te_{1/2}$ superconductor. The result of resistivity and magnetization measurements clearly show that title compound becomes a superconductor below ~14 K. $\rho(T)H$ measurements shows that the superconductivity transition temperature ($T_c$) barely decreases with applied field. This is unlike the Nb based BCS type, HTSC, Borides or even the recently invented FeAs superconductors. The high $H_{c2}(0)$ value render this superconductor as powerful competitor, which will be potentially useful in very high field applications.

**Acknowledgement**

The work is supported by Indo-Japan (DST-JSPS) bilateral exchange research program. VPSA further thanks NIMS for providing him with the MANA visiting scientist position for three months. Authors from NPL would like to thank Prof. Vikram Kumar (DNPL) for his constant encouragement. Anand Pal, Arpita Vajpayee and Monika Mudgel are thankful to CSIR for providing the financial support during their research.

**Figure Captions:**

Figure 1: The observed (red dots), calculated (solid line) and differences diffraction (bottom solid line) profiles at 300 K for $FeSe_{1/2}Te_{1/2}$.

Figure 2: Temperature dependence of magnetization under 10 Oe field measured in both field cooling (FC) and zero-field cooling (ZFC) conditions; Lower inset shows the complete M(H) loop at 2, 5 & 8 K and upper inset shows the first quadrant of the M(H) loop.

Figure 3: Temperature dependence of the resistivity of $FeSe_{1/2}Te_{1/2}$ measured in 0 kOe field; lower inset shows the temperature dependence of resistivity measured in fields up to 90 kOe for the same compound and upper inset shows the $H_{c2}$ vs T plots derived from measurements of resistivity against temperature and magnetic field [ρ(T)H plots].

Figure 4: Heat capacity variation with temperature ($C_p$-T) in zero-field and 70 kOe field; inset shows the enlarged view of the anomaly present in $C_p/T$ vs $T^2$.

Fig. 1

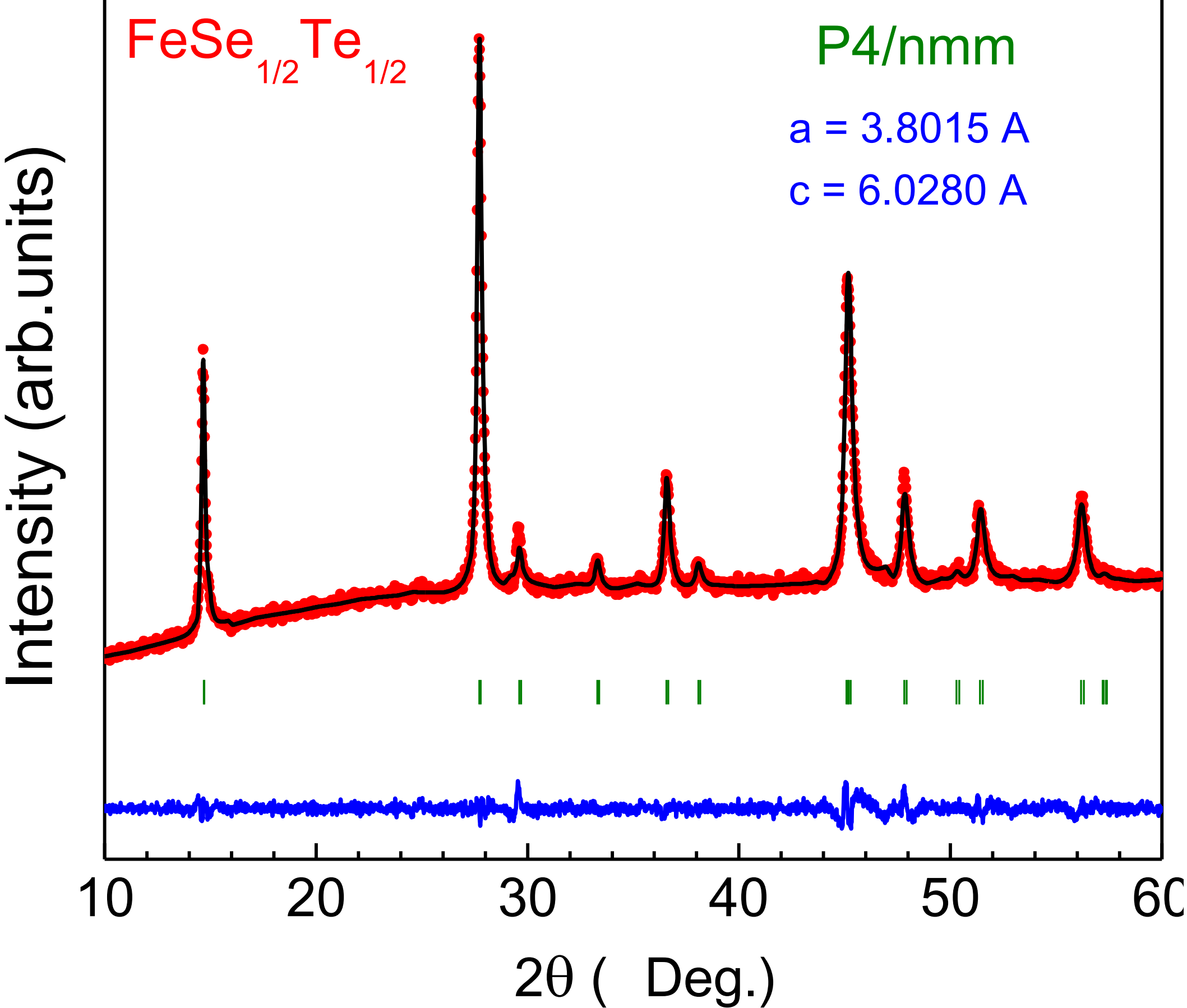

Fig. 2

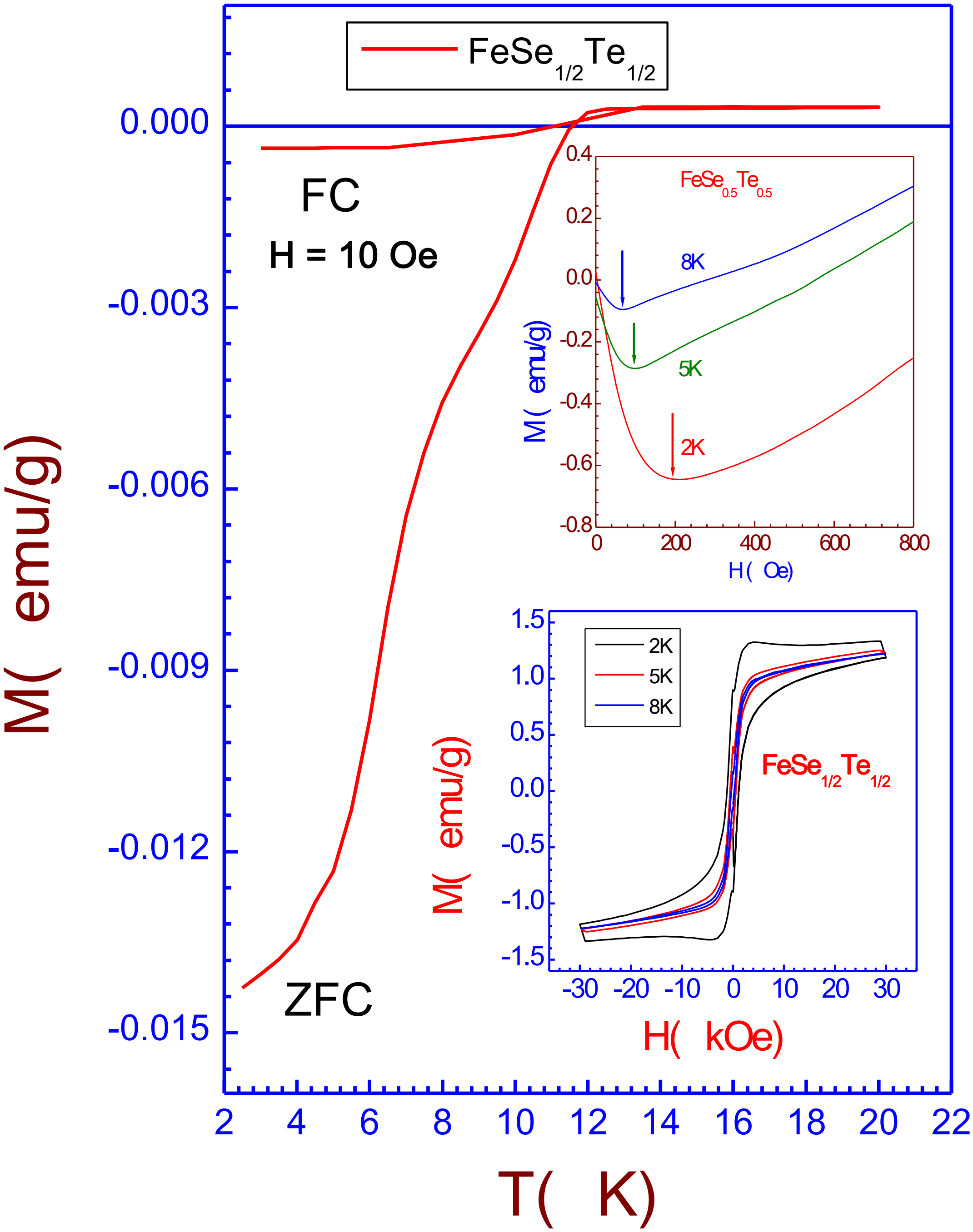

Fig. 3

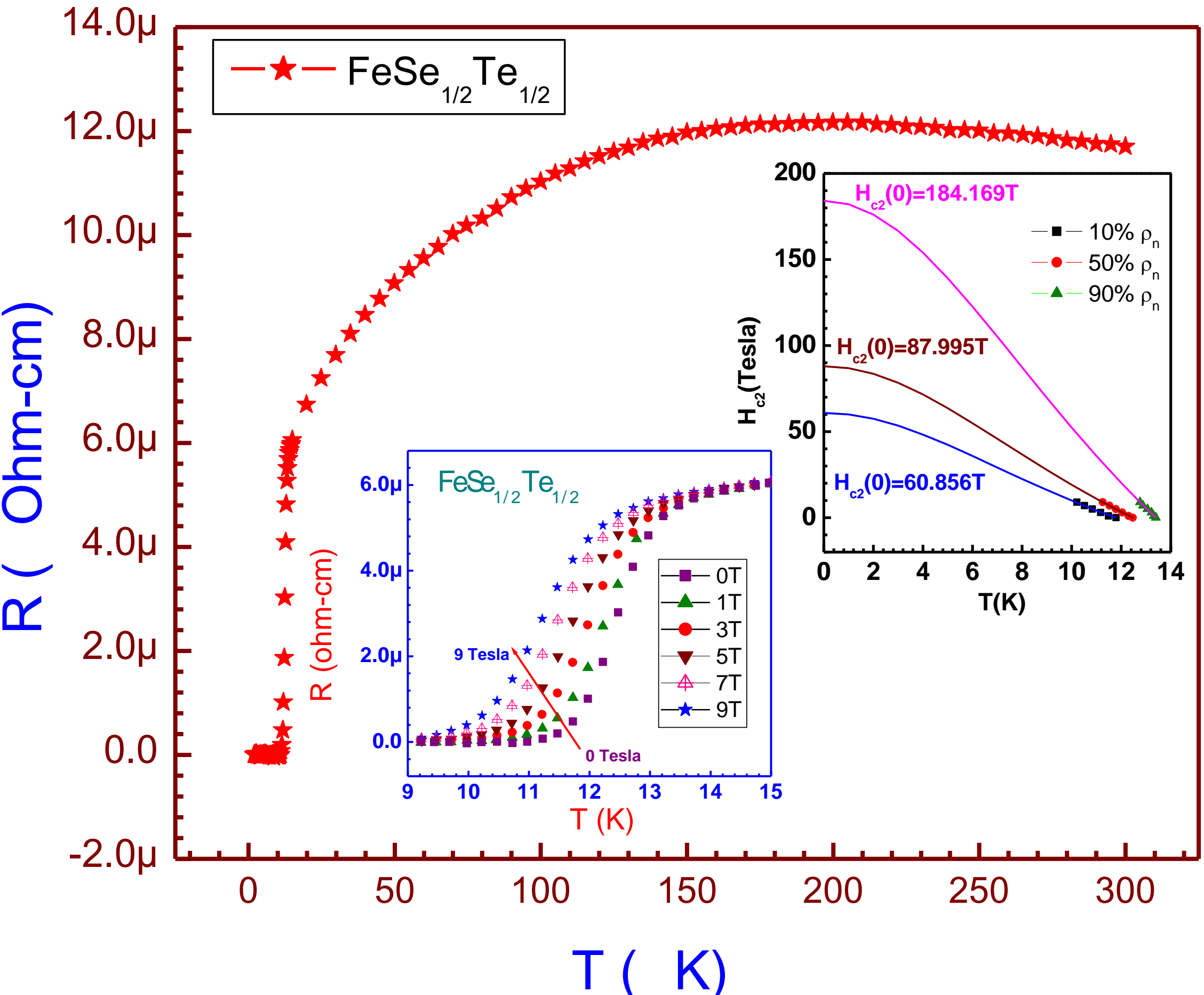

Fig. 4

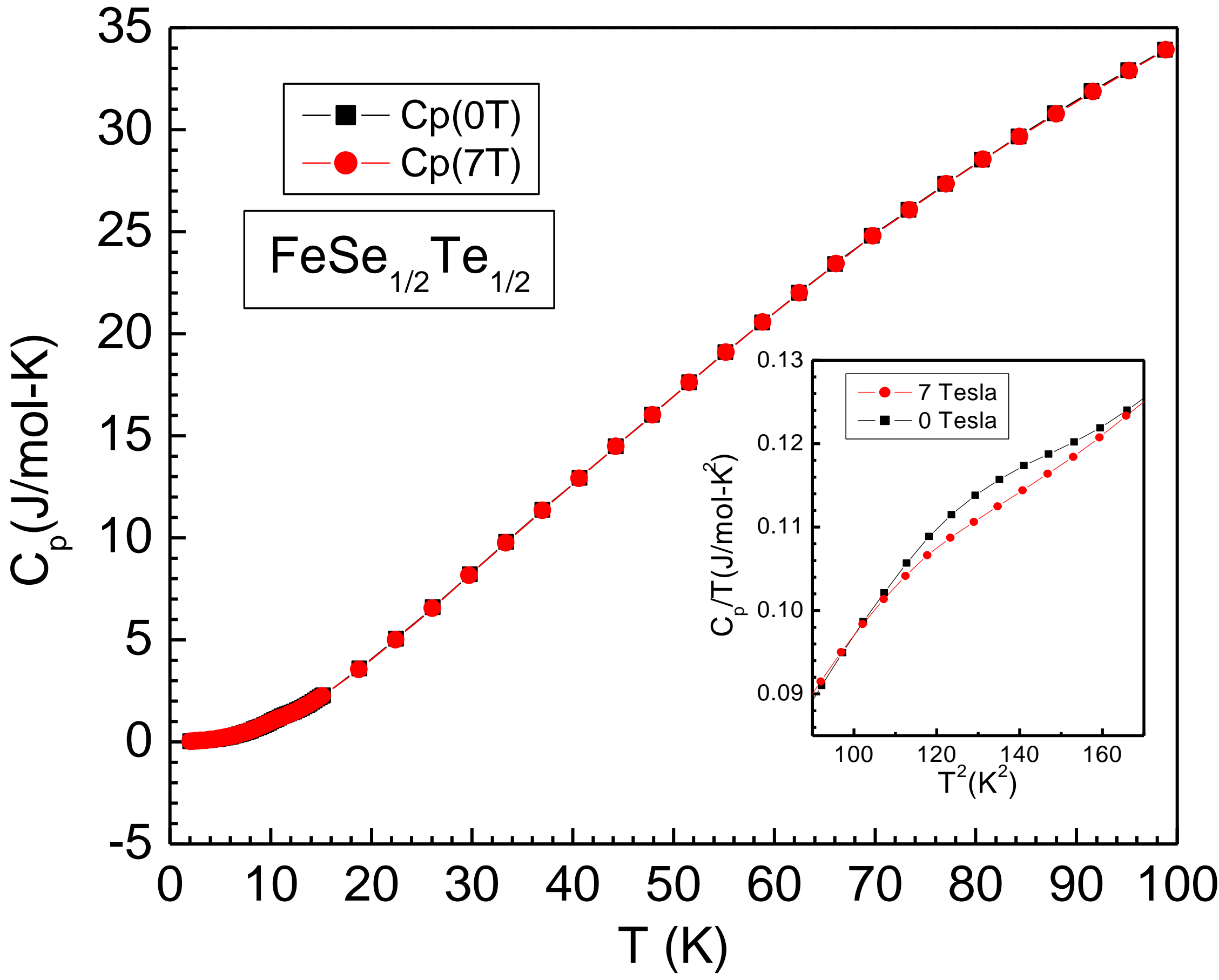